\documentclass[aps,pra,twocolumn,showpacs]{revtex4-1}
\usepackage{graphicx,color}
\usepackage{amsmath,amsfonts}
\usepackage{braket}
\usepackage[percent]{overpic}
\usepackage{hyperref}

\newcommand{\etal}{{\it et al.}~}

\newcommand{\Nv}{N_{\rm v}}

\begin{document}

\title{Decay of homogeneous two-dimensional quantum turbulence}

\author{Andrew W. Baggaley$^1$ and Carlo F. Barenghi$^1$}
\affiliation{$^1$Joint Quantum Centre Durham-Newcastle, 
School of Mathematics, Statistics and Physics, Newcastle University,
Newcastle upon Tyne, NE1 7RU, United Kingdom}
\date{\today}


\begin{abstract}
We numerically simulate the free decay of two-dimensional quantum turbulence
in a large, homogeneous Bose-Einstein condensate. The large number of vortices,
the uniformity of the density profile and the absence of 
boundaries (where vortices can drift out of the condensate) 
isolate the annihilation
of vortex-antivortex pairs as the only mechanism which reduces the number of
vortices, $\Nv$, during the turbulence decay. 
The results clearly reveal that vortex annihilations is a four-vortex 
process, confirming the decay law $\Nv \sim t^{-1/3}$ where $t$ is time,
which was inferred from
experiments with relatively few vortices in small harmonically trapped
condensates.
\end{abstract}

\maketitle
\section{Motivation}

Quantum turbulence (the chaotic motion of quantum vortices in 
superfluid helium\cite{Barenghi-PNAS} and cold gases\cite{Tsatsos}) 
has become a prototype problem of nonlinear statistical physics.  
The absence of viscosity and the nature of vorticity distinguish
quantum turbulence from ordinary turbulence: in quantum fluids infact,
vorticity is not a continuous field of arbitrary shape and strength
(as in ordinary fluids),
but is concentrated on the nodal points (in 2D) or lines 
(in 3D) of a complex wavefunction $\psi$. Around these points or lines
where $\psi=0$, the
phase of $\psi$ changes\cite{multi} by $2 \pi$. 
The large scale properties of quantum turbulence thus depend on the
interactions of discrete vortices, which induce effects such as
Kelvin waves\cite{Kivotides2001,Simula2008,Fonda2014,diLeoni2016}, 
vortex reconnections\cite{Zuccher2012,Serafini2016,Villois2017} and phonon 
emission\cite{Leadbeater2001,Leadbeater2003}. 
At temperatures
sufficiently close to the critical temperatures, 
the interaction of vortices with thermal 
excitations\cite{Jackson2009,Kim2016} induces 
friction effects \cite{BDV}.

In turbulence, the study of the free decay is fruitful because
it removes the arbitrariness of the forcing which is necessary to sustain
a statistical steady state.  
In 3D, experiments\cite{Walmsley2008,Zmeev2015} and numerical 
simulations \cite{Baggaley2012} of the decay of quantum turbulence 
in superfluid helium have revealed the existence of two turbulent regimes: a 
quasi-classical (or Kolmogorov) regime, which decays as $L(t) \sim t^{-3/2}$,
and an ultra-quantum (or Vinen) regime, which decays as $L(t) \sim t^{-1}$,
where the vortex line density $L$ (defined as the length of vortex
lines per unit volume) measures the turbulence's intensity. 
Physically, the Kolmogorov regime is characterized by a
cascade of kinetic energy from large to small eddies (similar to what
happens in ordinary turbulence), whereas the Vinen regime lacks a 
cascade and is more akin to a random flow \cite{SciRep2016}.
Recent studies of 3D turbulence in atomic condensates have
identified these two regimes \cite{Cidrim2017}, despite uncertainties 
due to the small number of vortices in the system
compared to liquid helium experiments.

In 2D, quantum turbulence takes the form of a chaotic configuration of
quantized point vortices. Since no direct vortex visualization is available
in superfluid helium films, all relevant 2D experiments have been performed
in trapped atomic Bose-Einstein condensates where vortices
can be easily imaged.
The 2D context has unique features (absent in 3D) associated to the possibility
of nonthermal fixed points\cite{Nowak2012}, 
an inverse energy cascade \cite{Reeves2013}
and the emergence of vortex clusters\cite{Billam2014,Simula2014}.
In this work we are concerned with a simpler question:
in analogy with 3D, what is the law governing the free decay of a random vortex 
configuration consisting of an equal number of positive and negative vortices ?
This question was experimentally addressed in a harmonically trapped condensate
by Kwon \etal \cite{Kwon2014}: they found that the time evolution of the
number of vortices, $\Nv(t)$ (the 2D equivalent of the vortex line density 
$L(t)$), is fairly well described by the logistic equation

\begin{equation}
\frac{d\Nv}{dt}=-\Gamma_1 \Nv - \Gamma_2 \Nv^2,
\label{eq:Shin}
\end{equation}

\noindent
In analogy with the kinetic theory of gases, Kwon \etal argued that the
rate coefficients $\Gamma_1$ and $\Gamma_2$ represent one-vortex and two-vortex
processes respectively:  the drift of vortices out of the condensate, and 
annihilations of vortex-antivortex pairs (the 2D analog of 3D reconnections). 
Stagg \etal \cite{Stagg2015} modelled numerically the experiment of
Kwon \etal, analyzed the results using Eq.~(\ref{eq:Shin}), and determined
that annihilations increase with temperature 
(see also \cite{Kim2016}).
Cidrim \etal \cite{Cidrim2016} attempted to generalize Eq.~(\ref{eq:Shin})
to the case of net polarization
$P=(\Nv^+-\Nv^-)/(\Nv^++\Nv^-) \neq 0$ 
(where $\Nv^+$ and $\Nv^-$ are the numbers
of positive and negative vortices respectively and $\Nv=\Nv^++\Nv^-$).
They noticed that the original interpretation of $\Gamma_1$ and
$\Gamma_2$ as one-vortex and two-vortex processes cannot be correct, as negative
values of $\Gamma_1$ were required to fit decays during which no vortices
visibly entered the condensate. Using different model equations for $\Nv^+$
and $\Nv^-$, they obtained a better fit to the observed decay. In the
case $P=0$ (corresponding to the experiment of Kwon \etal), 
the model of Cidrim \etal reduces to

\begin{equation}
\frac{d\Nv}{dt}=-\Gamma_1 \Nv^{3/2} - \Gamma_2 \Nv^4,
\label{eq:Cidrim}
\end{equation}

\noindent
where the $\Nv^{3/2}$ and $\Nv^4$ dependence
of the drift and the annihilation terms
were derived using physical arguments.
In particular, the quartic nature of the annihilation term 
in Eq.~(\ref{eq:Cidrim}) agrees
with the observation of Groszek \etal \cite{Groszek2016} that the annihilation
of a vortex anti-vortex pair is a four-vortex process,
not a two-vortex process (hence $\Nv^4$ rather than $\Nv^2$). Briefly, the
argument is the following. Without dissipation,
a vortex and anti-vortex alone would be a stable configuration which
travels at constant velocity. A third vortex is necessary 
to bring the two vortices together, destroying the circulation and creating 
a stable nonlinear wave; 
this wave, called `crescent-shaped' by Kwon \etal and
`vortexonium' by Groszek \etal, was identified as a soliton by
Nazarenko and Onorato \cite{Nazarenko2007,Stagg-JR}. 
The fourth vortex 
is necessary to destroy the nonlinear wave upon collision, radiating phonons 
away. Groszek \etal \cite{Groszek2016} also highlighted the role played
by the trapping potential; in particular, they found that vortex clustering 
is energetically less likely in harmonically trapped condensates compared 
to recently developed box-traps \cite{Gaunt2013,Navon2015}.

In contrast, in the presence of dissipation,
vortices of opposite circulation move towards one another 
and annihilate directly.  
Hence, it is natural to expect that in the presence of dissipation the decay of two-dimensional quantum turbulence follows a two-vortex
process. Indeed, our results will verify that this is the case.

Unfortunately the number $\Nv(t)$ of point vortices 
in the cited studies
is relatively small due to the constraints of current experimentally 
available condensates. The decay curves $\Nv(t)$ 
are therefore noisy, and it is difficult to determine with precision the 
exponents of the two effects - vortex drift and vortex annihilation.
Moreover, the drift of vortices out of the condensate
is likely to depend on the steepness of the trapping potential, which now
is not necessarily harmonic \cite{Gaunt2013,Navon2015}. 

To make progress towards understanding the law of 2D turbulence decay,
we concentrate on the annihilation process which here we study in the
absence of vortex drift by performing numerical simulations in a
uniform condensate without boundaries. In other words, we want to
determine accurately the exponent $k$ of the rate equation
\begin{equation}
\frac{d\Nv}{dt}=-\Gamma_1 \Nv^k,
\label{eq:k}
\end{equation}
\noindent
when the only mechanism responsible for decreasing 
$\Nv(t)$ is annihilations of vortex-antivortex pairs. For large
times, the solution of Eq.~(\ref{eq:k}) scales as $\Nv \sim t^{1/(1-k)}$, 
if $k>1$.  A precise measurement of the exponent $k$
will help future works to
determine the decay in finite-sized, non-uniform condensates,
where the decay depends also on vortices drifting out of the
boundaries.

\section{Model}

Our model is the 2D Gross-Pitaevskii equation (GPE) for an atomic 
condensate 

\begin{equation}
(i-\gamma) \hbar \frac{\partial \psi}{\partial t}=-\frac{\hbar^2}{2m} 
\left( \frac{\partial^2 \psi}{\partial x^2} +
\frac{\partial^2 \psi}{\partial y^2} \right) +
g \vert \psi \vert^2 \psi - \mu \psi,
\label{eq:GPE}
\end{equation}

\noindent
where $\psi(x,y,t)$ is the wavefunction, $m$ is the boson mass, $g$
is the interaction strength, $\mu$ is the chemical potential, 
$\hbar=h/(2 \pi)$ and
$h$ is Planck's constant. The phenomenogical dissipation 
coefficient $\gamma$ \cite{Tsubota2002}
is used in some our numerical simulations to mimic 
the interaction of the condensate with the thermal
cloud, in particular the loss of energy (i.e. the reduction in size)
of vortex-antivortex pairs.

Eq.~(\ref{eq:GPE}) is made dimensionless using the length scale
$\xi=\hbar/\sqrt{m \mu}$, the time scale $\hbar/\mu$ and the
density scale $\vert \psi \vert^2=\mu/g$, and solved
in the (dimensionless) periodic domain $-D \leq x,y \leq  D$ 
with $D=512\xi$. The large size of the domain (compared to the vortex
core size which is of the order of $\xi$) and the absence of
boundaries allow us to track the evolution and annihilations
of thousands of vortices, a number which is larger than in the typical
experiments and previous numerical simulations.
Space is discretized
 onto a $N=2048^2$ uniform cartesian mesh, spacial derivatives are approximated by a $6^{\rm th}$--order finite difference scheme and a $3^{\rm rd}$--order Runge-Kutta scheme is used for time evolution. 

The initial conditions of our simulations consist of a large number $\Nv$ of
vortices with approximately net zero polarization ($\Nv^+ \approx \Nv^-$). 
To create this condition modelling an experimentally feasible manner, 
we initialize the system with the 
non-equilibrium state  \cite{Berloff2002,Stagg2016}, 
$\psi (\mathbf{x}, 0) =  \sum_{k} a_k \exp(i\mathbf{k} \cdot \mathbf{x})$, where $\mathbf{k}=(k_x,k_y)$ is the wavevector, and the coefficients $a_k$ are uniform and the phases are distributed randomly.
By taking $k_x,k_y \in \mathbb{Z}$ we ensure our initial configuration satisfies the periodic boundaries we impose.
We perform three sets of simulations; without dissipation ($\gamma=0$) and at two different levels of dissipation, $\gamma=0.01$ \& $0.0025$. 
To ensure our results are independent of the initial conditions we make use 
of ensemble averaging and all results presented are averaged over 
simulations from 10 different initial non-equilibrium states.

\begin{figure*}
\begin{center}
    \includegraphics[width=0.485\textwidth]{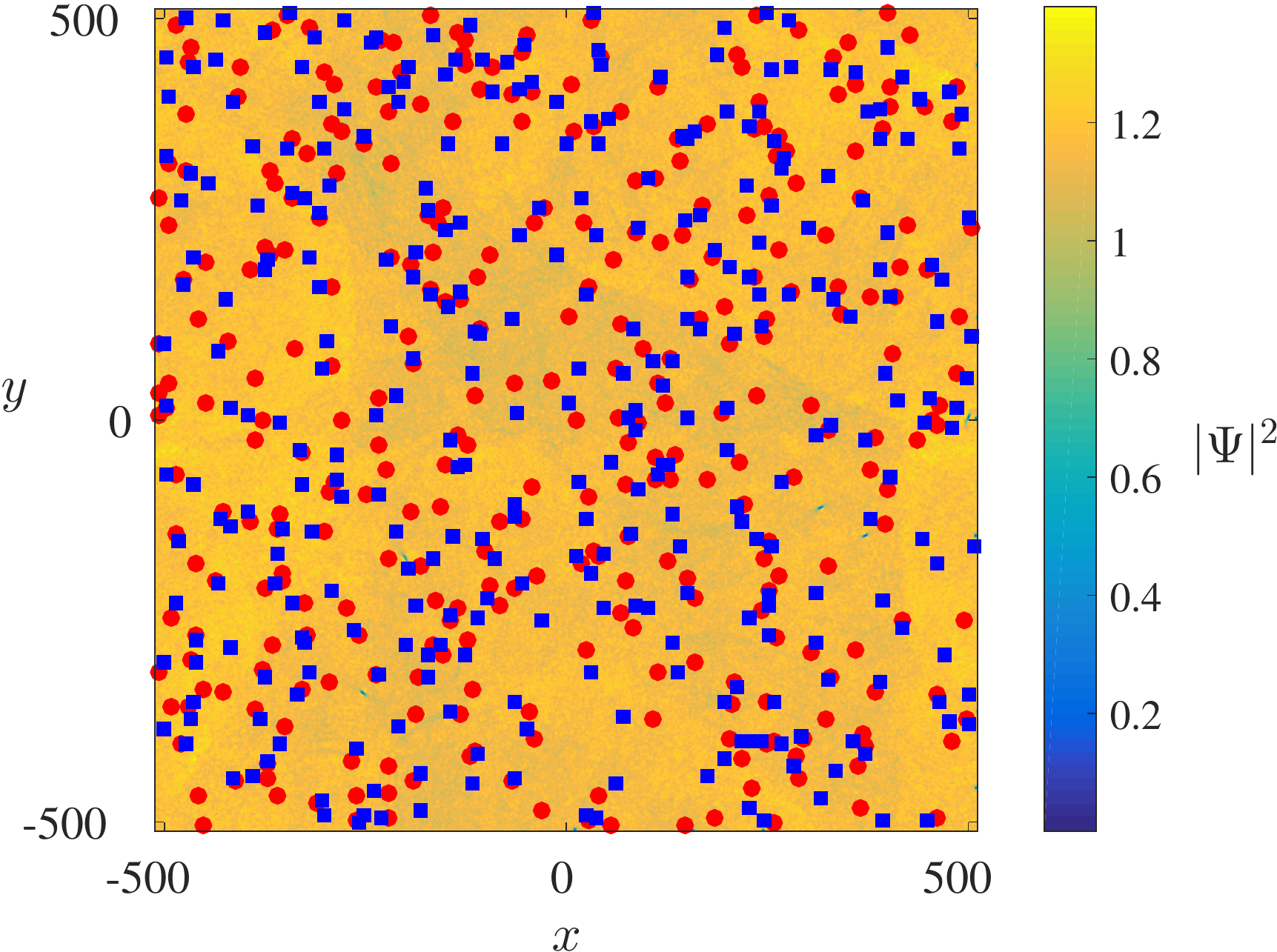} 
    \hfill
    \includegraphics[width=0.485\textwidth]{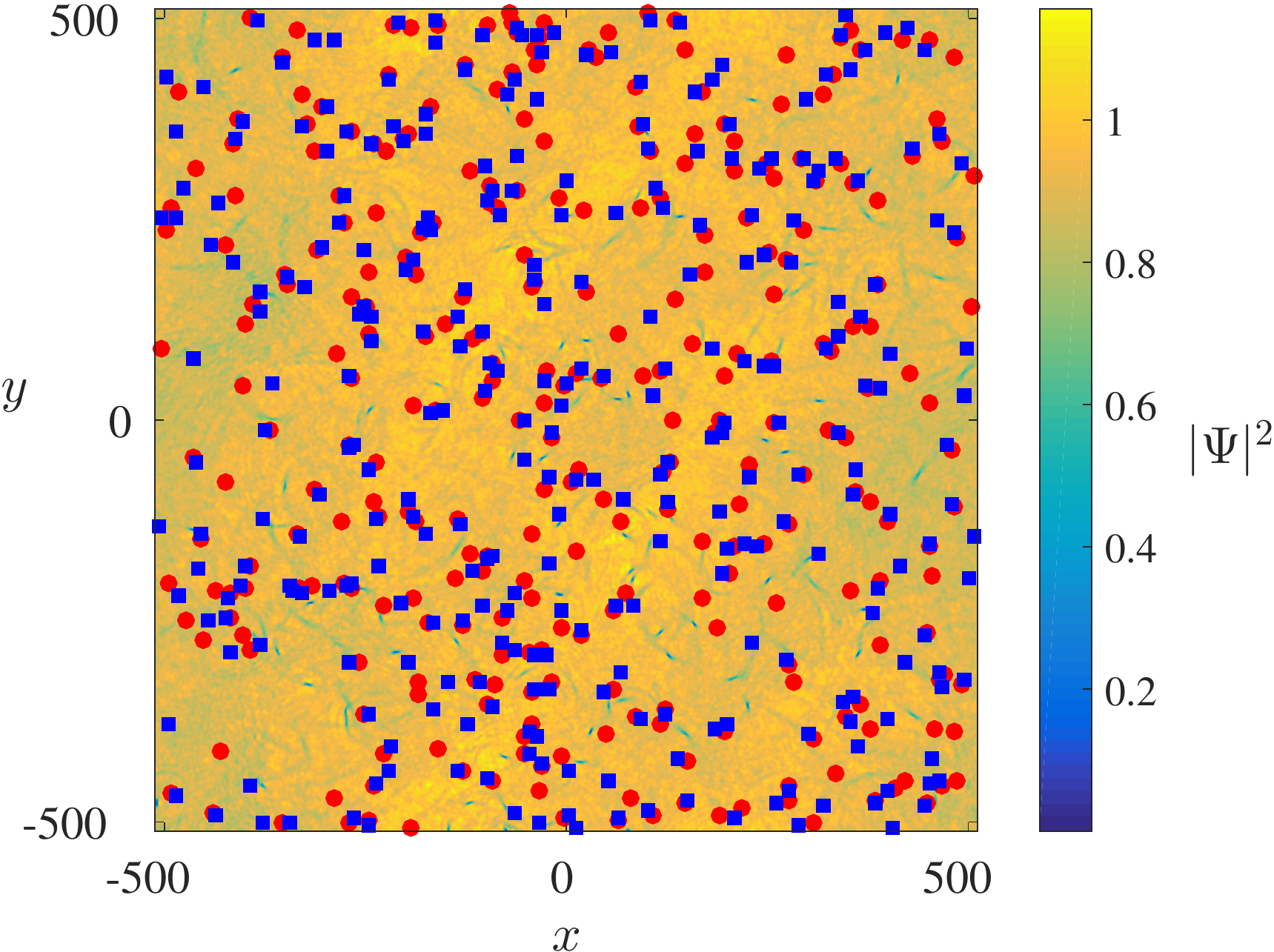}
\caption{(Color online).
Condensate's density $\vert \psi \vert^2$ vs $x,y$
during the decay of quantum turbulence without
dissipation ($\gamma=0$, left) and with dissipation
($\gamma=0.01$, right). 
Vortex locations are inferred by an algorithm which identifies
the $2\pi$ phase winding and the associated density depletion.
We mark the location of vortices with a positive circulation with a red circle and, those with a negative circulation using a blue square.}
\label{fig0}
\end{center}
\end{figure*}

During the time evolution we compute the total number of vortices based 
on a previously tested algorithm \cite{Stagg2015} which identifies 
locations where the condensate possesses a $2\pi$ winding of the phase,
and an associated density depletion,
see Fig.~\ref{fig0}.

\section{Results}

\begin{figure*}
\begin{center}
    \includegraphics[width=0.485\textwidth]{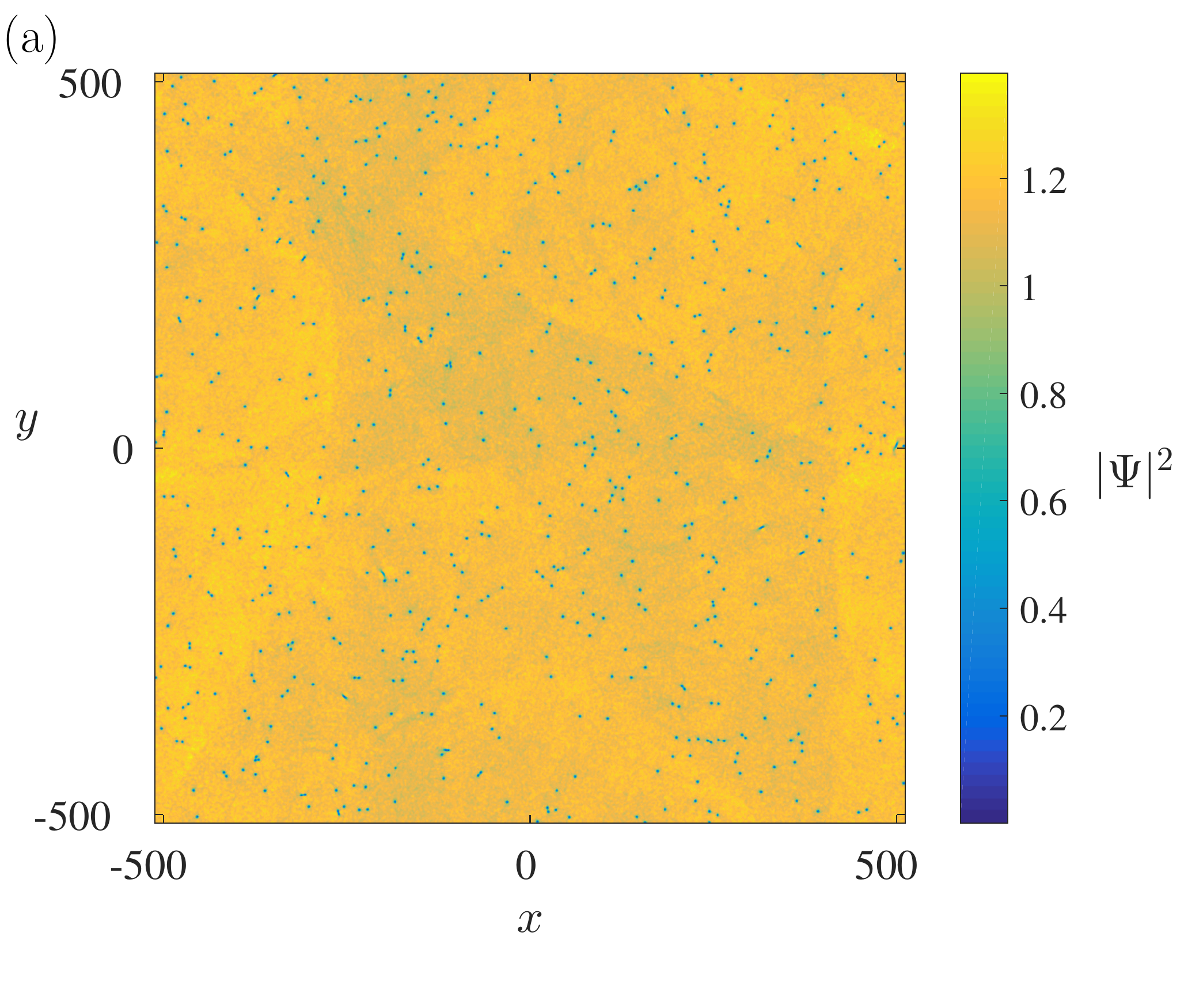} 
    \hfill
    \includegraphics[width=0.485\textwidth]{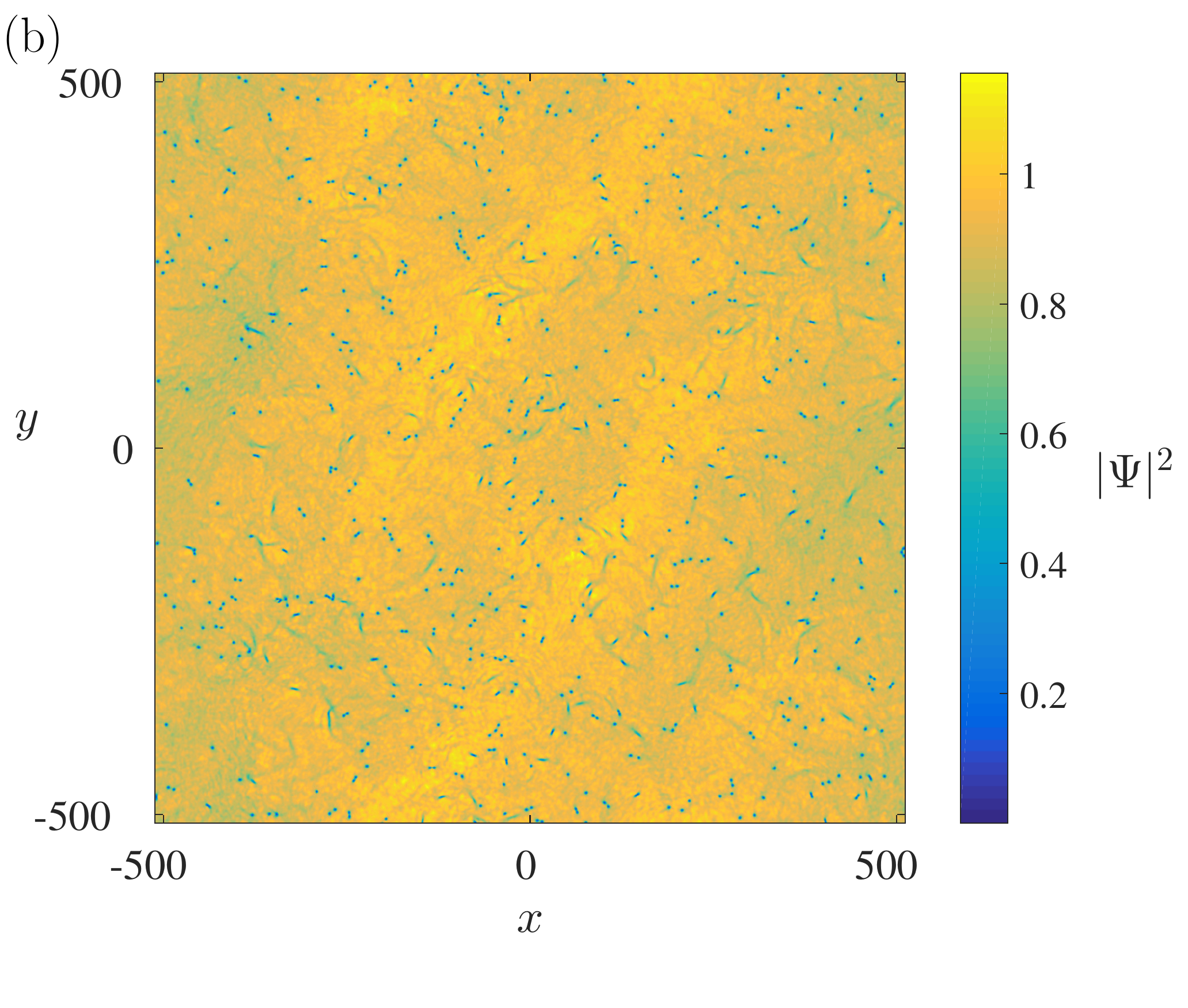}\\ 
       \includegraphics[width=0.485\textwidth]{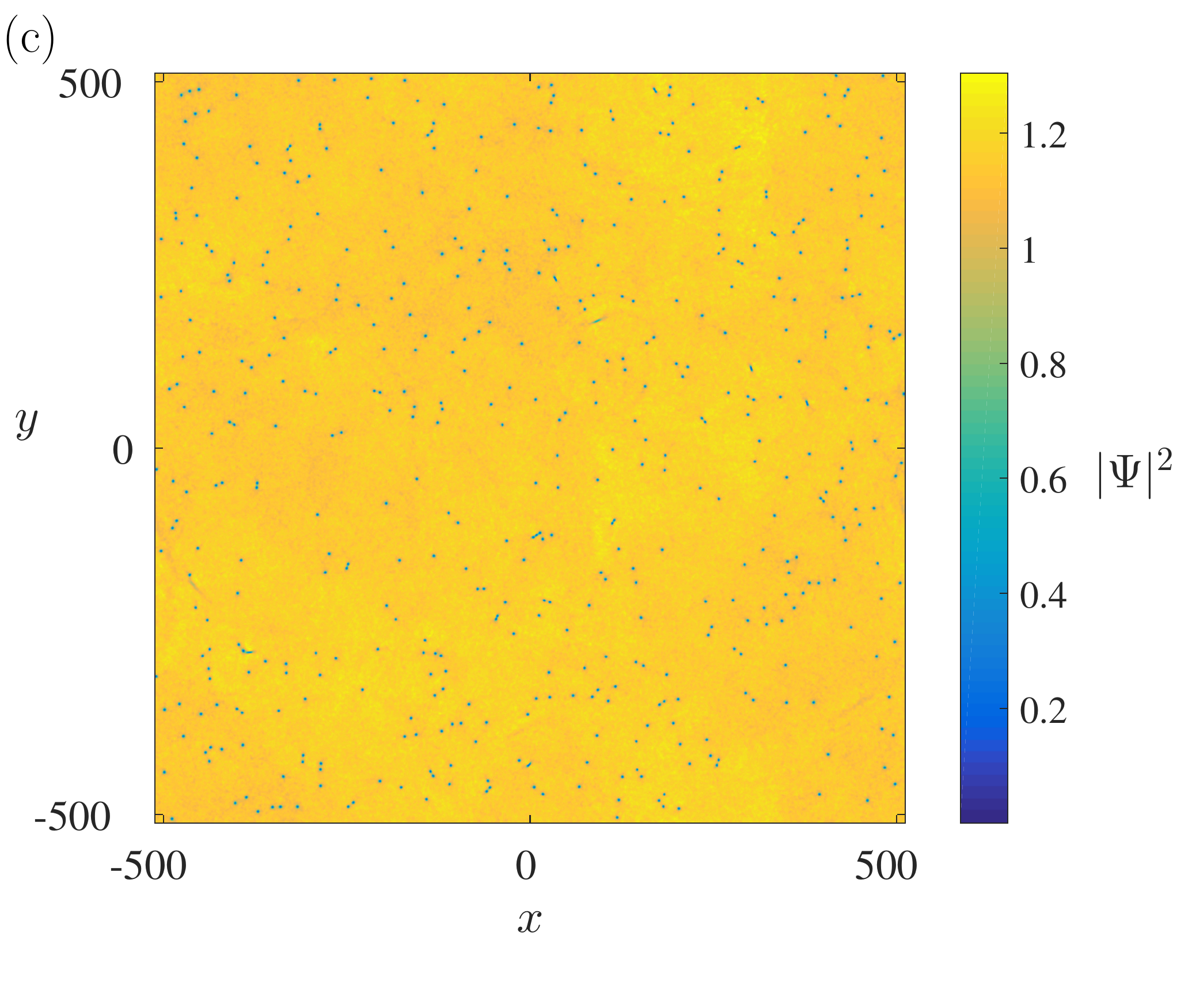} 
       \hfill
    \includegraphics[width=0.485\textwidth]{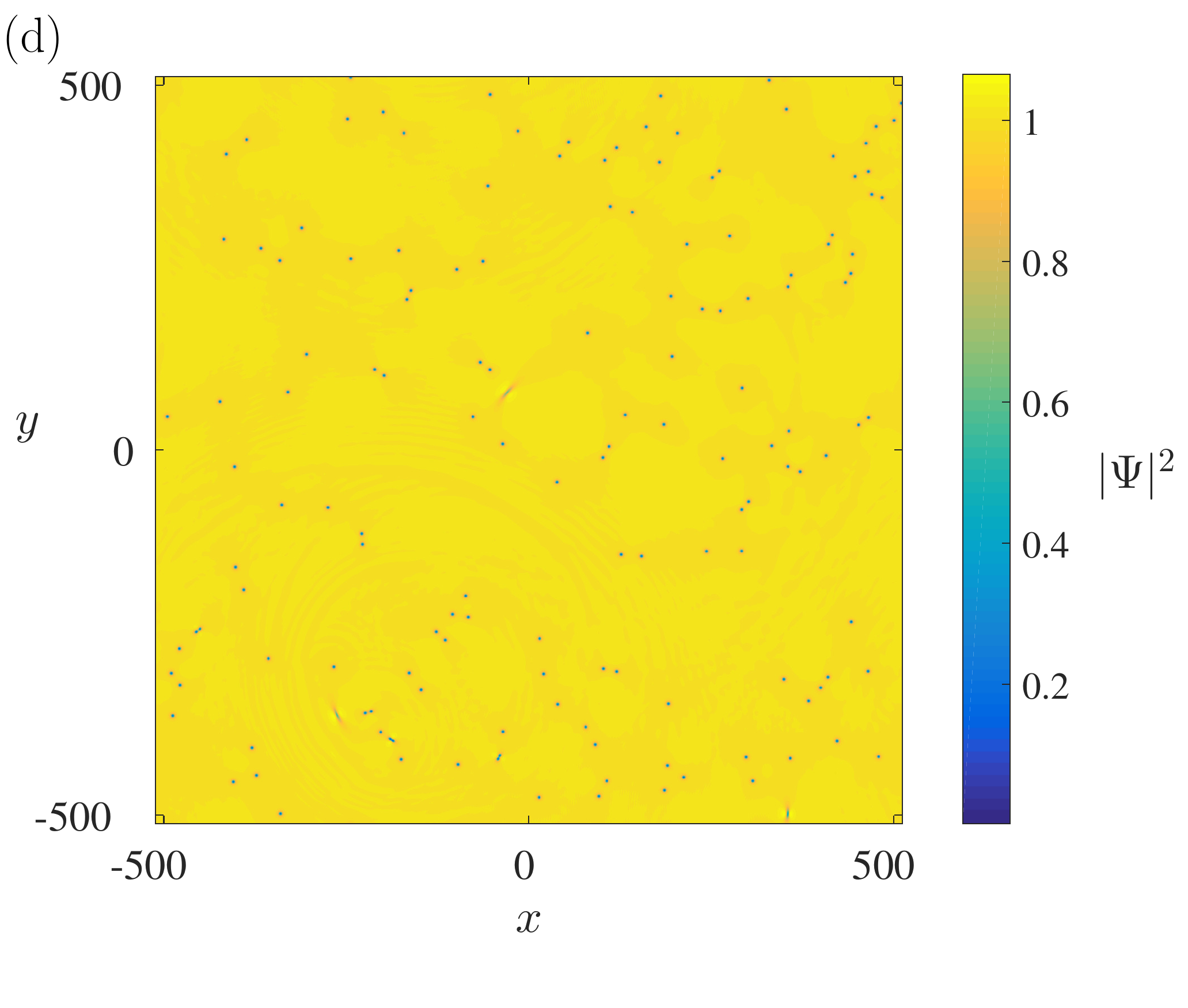}\\ 
       \includegraphics[width=0.485\textwidth]{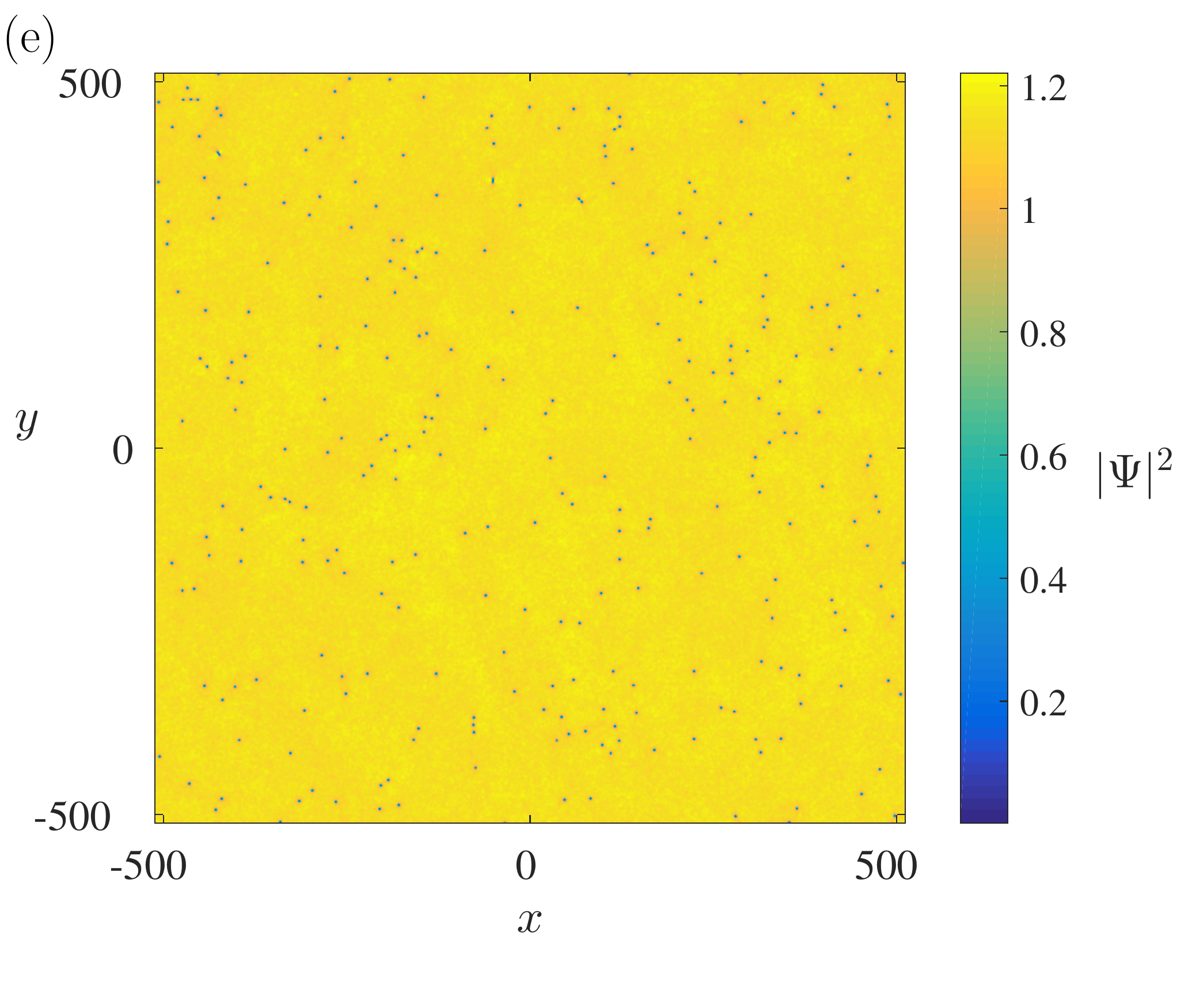} 
       \hfill
    \includegraphics[width=0.485\textwidth]{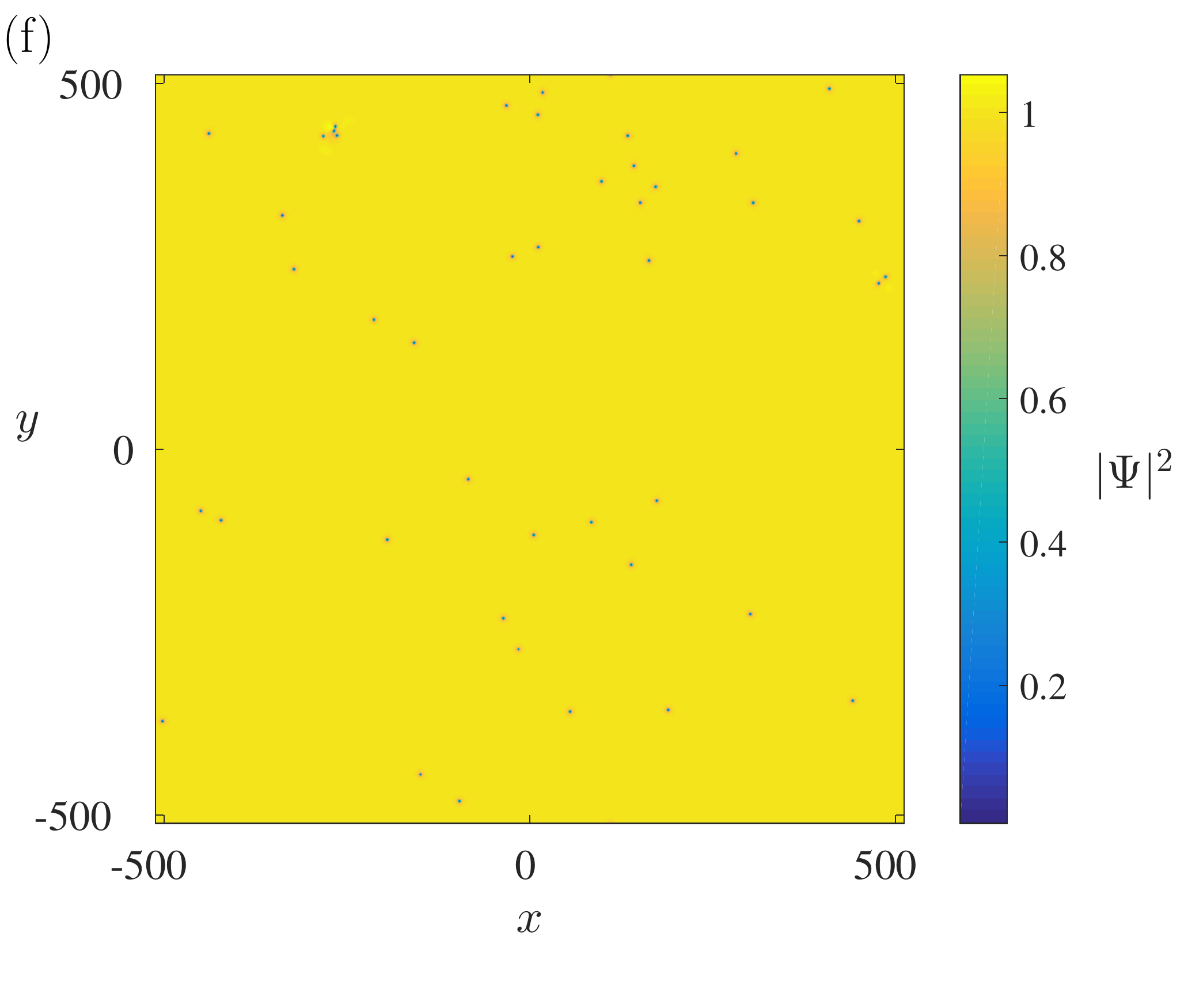}\\ 
\caption{(Color online).
Condensate's density $\vert \psi \vert^2$ vs $x,y$
during the decay of quantum turbulence without
dissipation ($\gamma=0$, left) and with dissipation ($\gamma=0.01$, right)
at times a) $t=7.5\times 10^3$ ; c) $t=2.5\times 10^4$; e) $t=1\times 10^5$, b) $t=1\times 10^3$ ; d) $t=1\times 10^4$; f) $t=5\times 10^4$.
The small holes in these density plots are the vortices.}
\label{fig1}
\end{center}
\end{figure*}

\begin{figure}
\begin{center}
\includegraphics[width=0.45\textwidth]{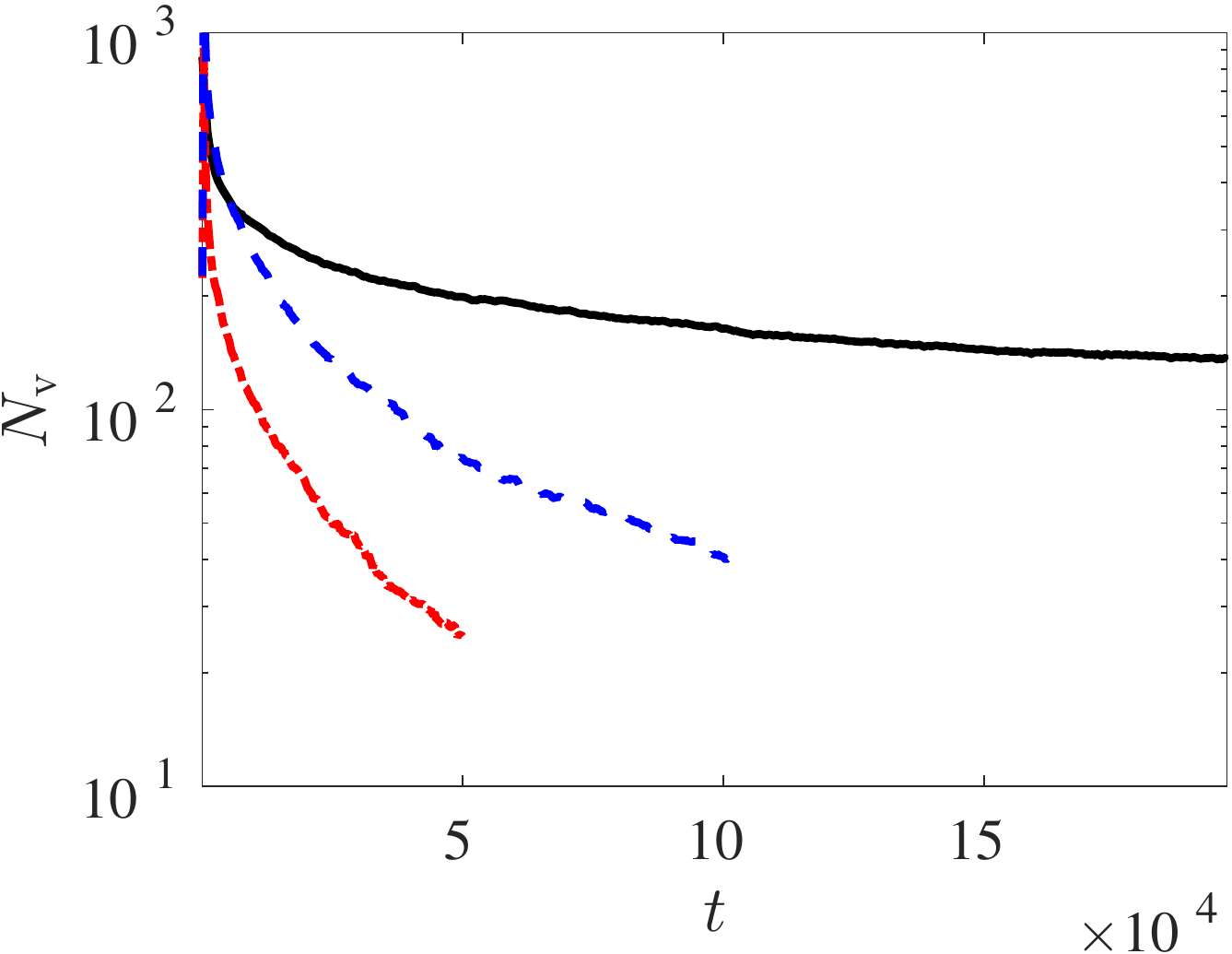} 
\caption{(Color online).
The total vortex number, $\Nv$, plotted vs time, $t$. 
The solid black curve, dashed blue curve and dot-dashed red curve correspond to 
ensemble-averaging 10 simulations without dissipation ($\gamma=0$)
and with dissipation ($\gamma=0.0025$ and $\gamma=0.01$) respectively. Notice the more rapid
decay induced by increasing dissipation.}
\label{fig2}
\end{center}
\end{figure}

Figure~\ref{fig1} displays the evolution of the condensate density 
on the $x,y$ plane at different 
times $t$ for the non-dissipative($\gamma=0$, left column) case
and a
dissipative ($\gamma=0.01$, right column) case. 
As vortices move chaotically
(accelerate) in each others' velocity fields, they radiate sound waves
\cite{Leadbeater2003}, turning
part of their kinetic energy into acoustic energy (phonons). Two vortices
of opposite signs which collide annihilate, radiating more sound
energy\cite{Leadbeater2001,Zuccher2012}. It is apparent from the figure that
dissipation damps out density oscillations and removes vortices
more quickly. The number of vortices $\Nv$ vs time $t$ (ensemble-averaged over
10 simulations) is displayed in Fig.~\ref{fig2} for both non-dissipative and 
dissipative cases. As expected, the decay of vortices is much faster 
in the presence of (larger) dissipation.

\begin{figure*}
\begin{center}
\includegraphics[width=0.32\textwidth]{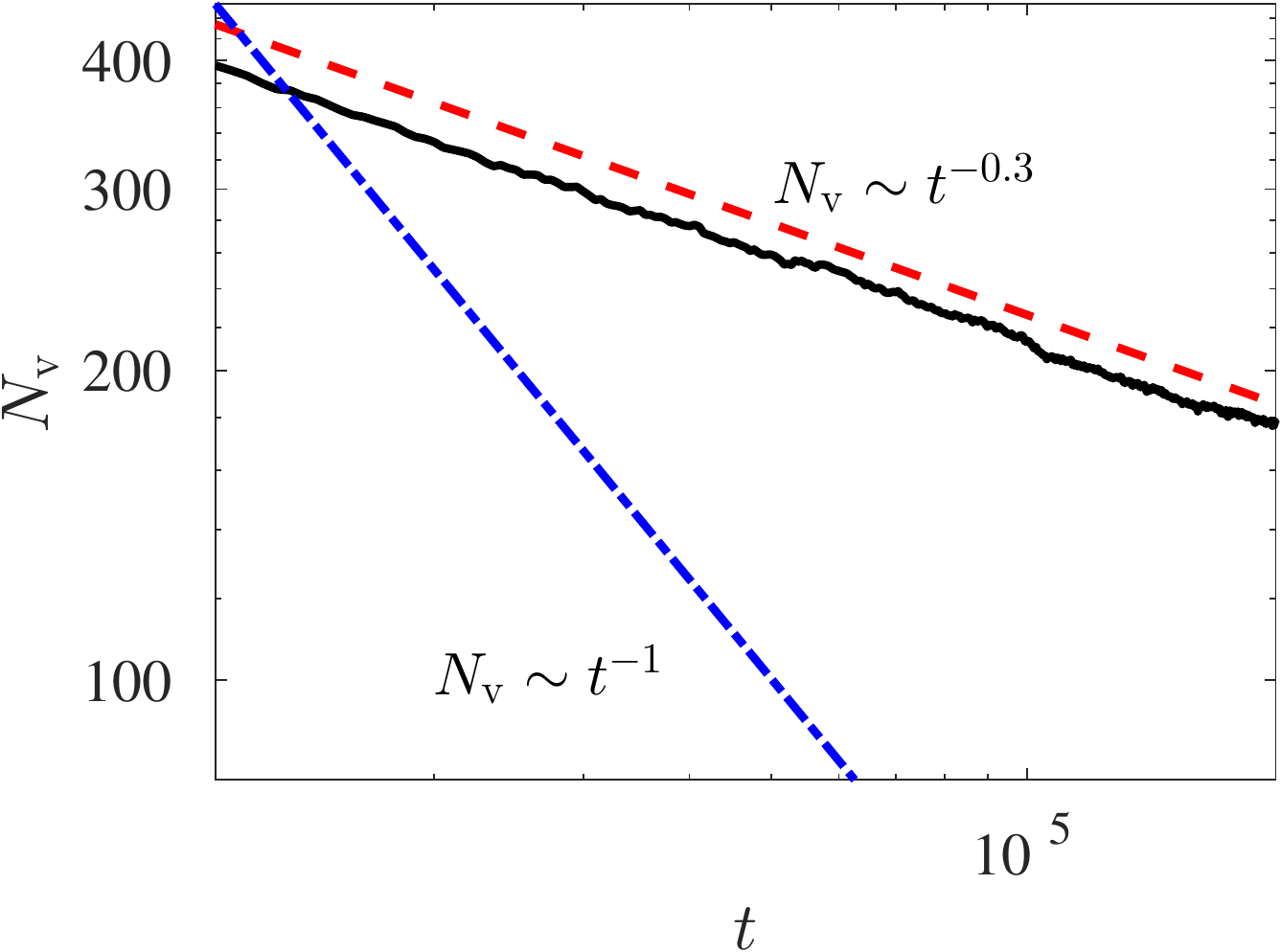} 
\includegraphics[width=0.32\textwidth]{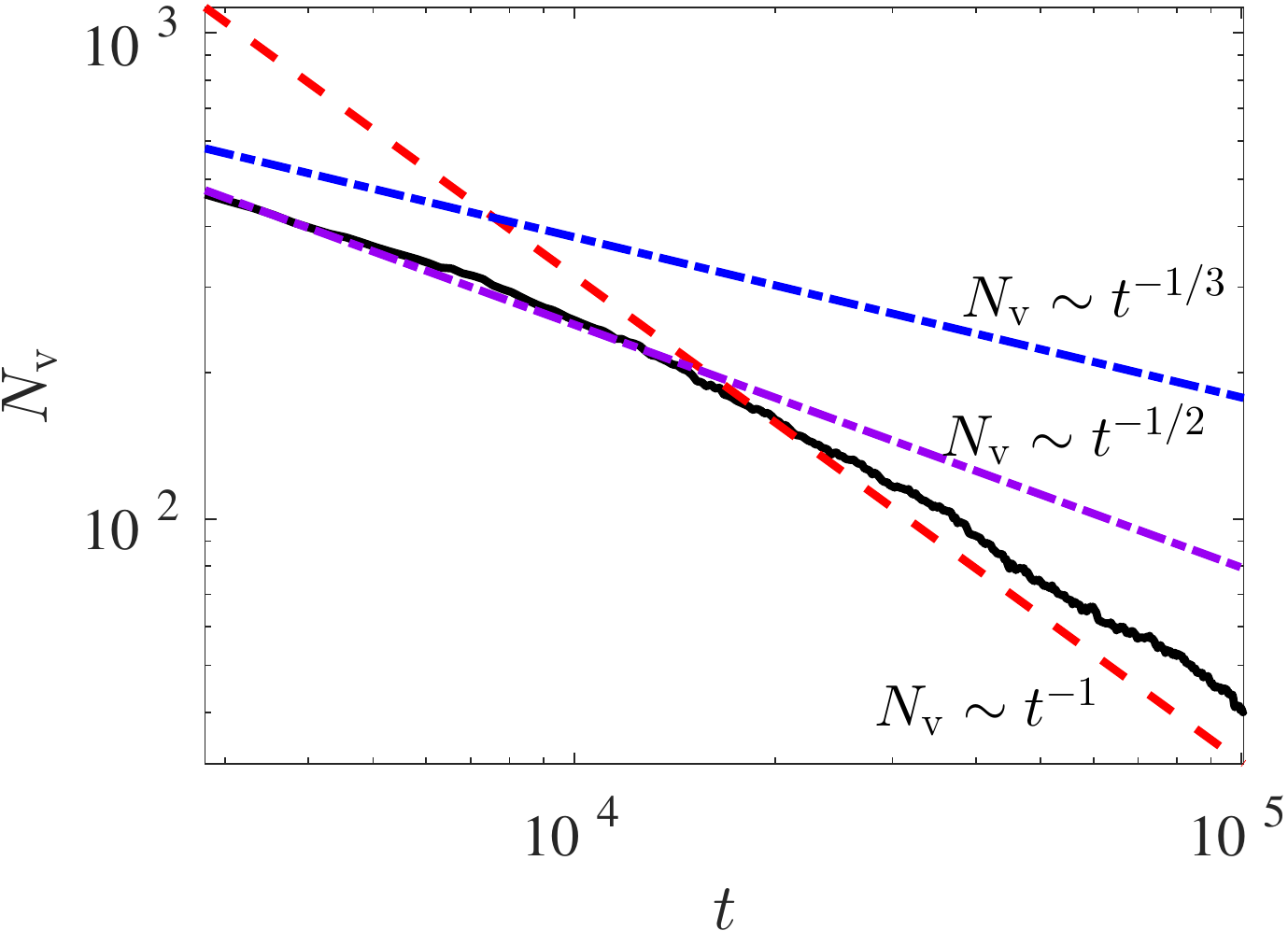}
\includegraphics[width=0.32\textwidth]{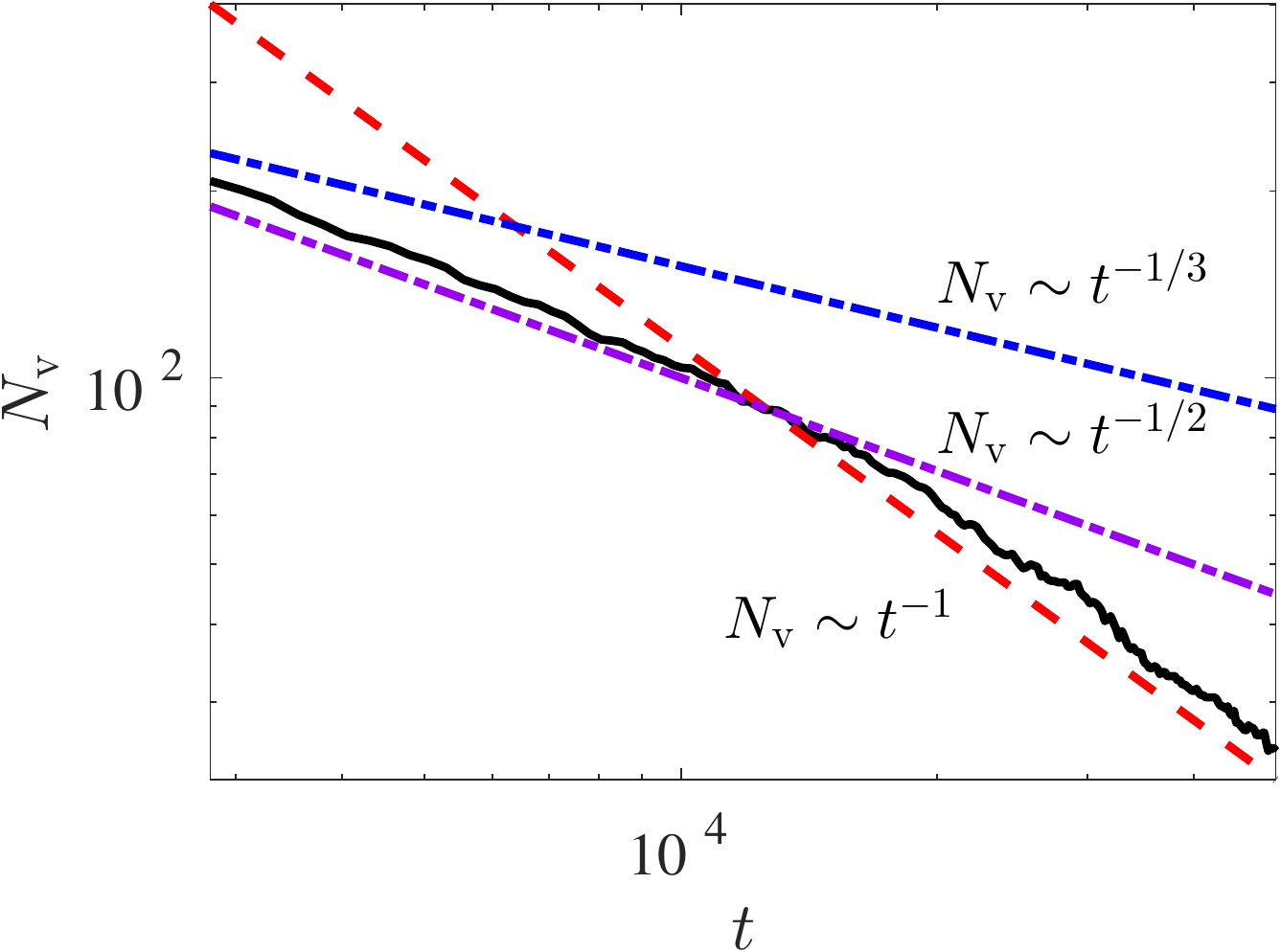} 
\caption{(Color online) Log log plots of the data presented in 
Fig.~\ref{fig2}, with corresponding best fits plotted as red dashed lines 
(position adjusted for clarity). The left panel corresponds to
the case $\gamma=0$, the central panel to $\gamma=0.0025$, and 
the right panel to $\gamma=0.01$. Alternative theoretical fits (discussed
in the text) are plotted as dot-dashed lines.}
\label{fig3}
\end{center}
\end{figure*}

Figure ~\ref{fig3} analyses the decay in a quantitative way.
The left panel of Fig.~\ref{fig3} shows that, in the absence of dissipation,
the vortex number decays as $\Nv(t) \sim t^{-0.3}$
in agreement with the $k=4$ scaling in Eq.~(\ref{eq:k}) of a
four-vortex process\cite{Cidrim2016,Groszek2016} which would yield
$\Nv \sim t^{-1/3}$ (red dashed line).  
The blue dot-dashed line of this panel
shows that the exponent $k=2$ of the two-vortex
process would not be a good fit.

The central and right panels 
of Fig.~\ref{fig3} show that, with $\gamma=0.0025$ \& $0.01$, 
the final part of the decay is steeper ($\Nv \sim t^{-1}$) and
more similar (particularly for $t > 2 \times 10^4$) 
to the prediction $\Nv \sim t^{-1}$ (red dashed line) of the 
two-vortex process. Clearly, the $N_v \sim t^{-0.3}$ decay
associated with the four-vortex process would not be a good fit at large
times.

We also observe that dissipation introduces a transient 
$\Nv \sim t^{-1/2}$ regime (clearly visible in the central and right panels)
before the final $N_v \sim t^{-1}$ regime is achieved.
This transient regime is the predicted outcome of a three-vortex process.
It seems reasonable to assume that early in the simulations,
when the vortex density is large, the annihilation of two vortices is 
predominantly induced by vortex dynamics (i.e. the presence of a
third vortex), and not by dissipation. The four-vortex scaling 
is not seen if the soliton that emerges from the annihilation is 
strongly damped by the dissipation. However once the vortex density 
becomes sufficiently small ($\Nv<100$ in these simulations),
the dissipation becomes the dominant mechanism which brings vortices 
together and annihilates them, hence the two-vortex scaling emerges.

\section{Conclusion}

We have performed numerical simulations of the free decay of 2D vortex
configurations, which initially contain thousands of vortices. 
The very
large homogeneous condensate and the absence of boundary effects has clearly
confirmed that vortex annihilation is a four-vortex process which is
described by the rate equation $d\Nv/dt=-\Gamma_1 \Nv^4$ proposed
by Cidrim \etal \cite{Cidrim2016} and Groszek \etal \cite{Groszek2016}.
The presence of dissipation adds additional complexity.
Initially the decay follows the three-vortex rate equation $d\Nv/dt=-\Gamma_1 \Nv^3$, as
dissipation eliminates the need for a fourth vortex to dissipation the resulting soliton.
However at small vortex densities dissipation brings the vortex and the closest antivortex
together without the need of the presence of other vortices, and the late-time decay
is faster, in qualitative agreeement with the rate equation
$d\Nv/dt=-\Gamma_1 \Nv^2$ first proposed by Kwon \etal \cite{Kwon2014}.

Having established the contribution to the turbulence decay
arising from the annihilation of vortices with antivortices,
it will be easier in future experiments to find the
contribution from vortices drifting out of the condensate 
(an effect which
likely depends
on the steepness of the confining potential).

\section{Acknowledgments}

This work was supported by EPSRC grant number EP/R005192/1.

\begin{thebibliography}{10}
\bibitem{multi}
More generally, the phase change is $2 \pi q$ where $q$ is an integer,
but vortices with $q>1$ are unstable.
\end{thebibliography}
\end{document}